\documentstyle[12pt,epic]{article}
\if@twoside \oddsidemargin 21pt \evensidemargin 59pt \marginparwidth 85pt
\else \oddsidemargin 0pt \evensidemargin 0pt
\fi
\newcounter{saveeqn}
\newcommand{\alpheqn}[1]{\refstepcounter{equation}\label{#1}%
\setcounter{saveeqn}{\value{equation}}%
\setcounter{equation}{0}%
\renewcommand{\theequation}
{\mbox{\arabic{saveeqn}\alph{equation}}}}
\newcommand{\reseteqn}{\setcounter{equation}{\value{saveeqn}}%
\renewcommand{\theequation}{\arabic{equation}}}
\newcounter{formel4}
\newcounter{formel5}

\begin{document}
                                                                  
\title{A Fluid Dynamic Model for the Movement of Pedestrians}

\author{Dirk Helbing\\ 
\small II. Institute for Theoretical Physics, University of Stuttgart,
Pfaffenwaldring 57/III\\
\small W-7000 Stuttgart 80, Germany}
\maketitle

\begin{abstract}
A kind of fluid dynamic description for 
the collective movement of
pedestrians is developed on the basis of a {\sc Boltzmann}-like
gaskinetic model. The differences between 
these pedestrian specific equations and those for
ordinary fluids are worked out, for example concerning the mechanism of
relaxation to equilibrium, the role of ``pressure'', the special influence
of internal friction and the origin of ``temperature''.
Some interesting results are derived that can
be compared to real situations, for example 
the development of walking lanes and of
pedestrian jams, the propagation of waves, and the behavior on a dance 
floor. Possible applications of the model to town- and traffic-planning
are outlined. 
\end{abstract}

\section{Introduction}

Former publications on the behavior of pedestrians have been mostly 
empirical (often in the sense of regression analyses)
and had the intention to allow planning of efficient
traffic \cite{Verbess,Verkehrsplanung,Planung}. 
Meanwhile there also exist theoretical approaches to pedestrian movement 
\cite{Borg1,Borg2,Borg3,Helbing1,Henderson,Push,San,Scott}. 
However, most theoretical work has been done in the related
topic of automobile traffic (see, for example, 
\cite{Alberti,Verkehr,stoch,Pav,Prig,Pri}). Especially, there have been
developed some {\sc Boltzmann}-like (gaskinetic) approaches
\cite{Alberti,Pav,Prig}. 
\par
As far as pedestrian movement is concerned, the author has observed that
footprints of pedestrian crowds
in the snow or quick-motion pictures of pedestrians look
similar to streamlines of fluids. It is the object of this paper to give
a suitable explanation of the fluid dynamic properties of pedestrian crowds.
{\sc Henderson} was the first, who compared gaskinetic
and fluid dynamic models to empirical data of pedestrian crowds
\cite{Henderson,Rennen,Frauen,Schiefe}. 
However, his work started from the conventional theory for ordinary
fluids, and assumed a
{\sl conservation of momentum and energy} to be fulfilled. 
In contrast to {\sc Henderson}'s approach, this article develops
a {\sl special theory for pedestrians}---without making use of the
unrealistic conservation assumptions. 
\par
We shall proceed in the following way: First the pedestrians will be
distinguished into groups of different types $\mu$ of motion, normally
representing different intended directions of walking.
At a time $t$ the pedestrians of each type $\mu$ of motion can
be characterized by several quantities like their place
$\vec{x}$, their velocity $\vec{v}_\mu$ and their intended velocity
$\vec{v}_\mu^0$ (that means the velocity they {\sl want} to walk with). 
So, in a given area $A$ can be found a 
density $\hat{\rho}_\mu
(\vec{x},\vec{v}_\mu,\vec{v}_\mu^0,t)$ of pedestrians 
having a special type
of motion $\mu$ and showing approximately the 
quantities $\vec{x}$, $\vec{v}_\mu$, $\vec{v}_\mu^0$ at time $t$.
For the densities $\hat{\rho}_\mu
(\vec{x},\vec{v}_\mu,\vec{v}_\mu^0,t)$ equations of motion can be set up
(section 2). From these equations we shall derive 
coupled
differential equations for the spatial density $\langle \rho_\mu \rangle$
of pedestrians, their mean velocity $\langle v_\mu \rangle$ and
velocity variance $\langle (\delta v_{\mu,i})^2 \rangle$ (section 3). 
The resulting
equations show many similarities to the equations for ordinary fluids, but
they contain some additional terms which take into account pedestrian
intentions and interactions
(sections 3.1, 4.1, 5.1 and 6). In section 4
we shall treat equilibrium situations and
the propagation of density waves. However, in {\sl non}equilibrium
situations the final adaptation time to local equilibrium gives rise
to internal friction (viscosity) and other additional terms (section 5). 
Effects of interactions (that means of avoiding maneuvers) between pedestrians 
will be discussed in section 6.
These effects will lead to some 
conclusions applicable to town- and traffic-planning
(section 7).
\par
Readers who are not interested in the mathematical aspects may skip the
formulas in the following. However, the mathematical results are important
for analytical, computational or empirical evaluations.

\section{Gaskinetic equations}

Pedestrians can be distinguished into different {\sl types}
$\mu$ {\sl of motion}, for example, by their different {\sl intended 
directions} $\vec{e}_\mu
:= \vec{v}_\mu^0 / \| \vec{v}_\mu^0 \|$
of motion (normally 2 opposite directions, at crossings 4 directions).
More exactly speaking, a pedestrian shall 
belong to a type $\mu$ of motion if it
wants to walk with an {\sl intended velocity} 
\begin{displaymath}
 \vec{v}^0 \in {\cal N}_\mu \, ,
\end{displaymath}
where 
\begin{displaymath}
 {\cal N}_\mu := \{ \vec{v}_\mu^0 \} 
\end{displaymath}
is one of several disjoint and complementary sets.
A type $\mu$ of motion still contains pedestrians with a variety of intended
velocities $\vec{v}_\mu^0$, but the advantage resulting from
a suitable choice of the sets ${\cal N}_\mu$ is to get
approximately {\sl unimodal} densities 
$ \hat{\rho}_\mu(\vec{x},\vec{v}_\mu,\vec{v}_\mu^0,t)$
and therefore to obtain appropriate mean value equations 
(see section \ref{mac}).
$ \hat{\rho}_\mu(\vec{x},\vec{v}_\mu,\vec{v}_\mu^0,t)$
describes the number $N_\mu$ of pedestrians of type
$\mu$ within an area $A=A(\vec{x})$ around place $\vec{x}$
having approximately the {\sl intended} velocity $\vec{v}_\mu^0$, but
approximately the {\sl actual} velocity $\vec{v}_\mu$. More exactly,
$\hat{\rho}_\mu$ is defined by
\begin{equation}
 \hat{\rho}_\mu(\vec{x},\vec{v}_\mu,\vec{v}_\mu^0,t)
\equiv \hat{\rho}_\mu(\vec{x},\vec{u}_\mu,t)
:= \frac{N_\mu({\cal U}(\vec{x})\times {\cal V}(\vec{u}_\mu),t)}{A\cdot V}\, ,
\label{density}
\end{equation}
where $N_\mu$ is the number of pedestrians of type $\mu$ 
which are at time $t$ in a {\sl state}
\begin{displaymath}
 (\vec{x}',\vec{u}'_\mu) \in {\cal U}(\vec{x}) \times {\cal V}(\vec{u}_\mu)
\end{displaymath}
out of the neighbourhood ${\cal U}(\vec{x}) \times {\cal V}(\vec{u}_\mu)$
of $\vec{x}$ and
\begin{displaymath}
\vec{u}_\mu := (\vec{v}_\mu,\vec{v}_\mu^0) \, .
\end{displaymath}
State $(\vec{x},\vec{u}_\mu)$ 
is an abbreviation for the property
\begin{displaymath}
(\vec{x},\vec{u}_\mu) :=  (\vec{x},\vec{v}_\mu,\vec{v}_\mu^0) \, ,
\end{displaymath}
that an individual is at place $\vec{x}$ and
wants to walk with the intended velocity
$\vec{v}_\mu^0$, but in fact walks with velocity $\vec{v}_\mu$.
\begin{equation}
{\cal U}(\vec{x}) :=
\{ \vec{x}^* \in {\cal M} : \| \vec{x}^* -\vec{x} \|_l
\le r \} \label{err}
\end{equation}
is a neighbourhood around the place $\vec{x}$ 
and belongs to the domain $\cal M$, which represents all {\sl accessible} 
(or public) places $\vec{x}$. $A=A(\vec{x})$ denotes the area of
${\cal U}(\vec{x})$. Similarly,
\begin{displaymath}
{\cal V}(\vec{u}_\mu) :=
\{ \vec{u}_\mu^* = (\vec{v}_\mu^*,\vec{v}_\mu^{0*})
: \| \vec{u}_\mu^* -\vec{u}_\mu \|_l
\le s, \vec{v}_\mu^0 \in {\cal N}_\mu \}  
\end{displaymath}
is a neighbourhood of $\vec{u}_\mu := (\vec{v}_\mu,\vec{v}_\mu^0)$
with a volume $V=V(\vec{u}_\mu)$.
\par
We shall now establish 
a set of {\sl continuity equations},
which are similar to the {\sl ansatz} of
{\sc Alberti} and {\sc Belli} \cite{Alberti}: 
\begin{eqnarray}
\frac{d\hat{\rho}_\mu}{dt} &\equiv&
\frac{\partial \hat{\rho}_\mu}{\partial t} 
+ \nabla_{\vec{x}}\, (\hat{\rho}_\mu \vec{v}_\mu) 
+ \nabla_{\vec{v}_\mu} \left(\hat{\rho}_\mu 
\frac{\vec{f}_\mu}{m_\mu}\right)
+ \nabla_{\vec{v}_\mu^0} \left(\hat{\rho}_\mu \dot{\vec{v}}{}_\mu^0 \right)
\nonumber \\
&:=& 
\frac{\hat{\rho}_\mu^0-\hat{\rho}_\mu}{\tau_\mu} 
+ \sum_\nu \hat{S}_{\mu \nu} + \sum_\nu \hat{C}_{\mu \nu} 
+ \hat{q}_\mu \, . \label{kont}
\end{eqnarray}
These equations can be interpreted as {\sl gaskinetic equations}
(see chapters 2.4, 2.7
of Ref. \cite{Kontin2}, and \S 3 of Ref. \cite{Kontin1}).
$m_\mu$ denotes the average 
mass of pedestrians belonging to type $\mu$ of
motion. Apart from special situations it will not depend on $\mu$,
that means $m_\mu \approx m_\nu$.
The forces $\vec{f}_\mu := m_\mu \dot{\vec{v}}_\mu$ can often
be neglected. However, they may be locally varying functions, depending
on the attraction of the places $\vec{x}$. The temporal change
$\dot{\vec{v}}{}_\mu^0$ of the intended velocity $\vec{v}_\mu^0$
can in principal be a function of place $\vec{x}$ and time $t$, but
it is normally a small 
quantity ($\dot{\vec{v}}{}_\mu^0 \approx \vec{0}$). Otherwise, 
the type $\mu$ of motion is changed.
\par
According to (\ref{kont}) the change of the density $\hat{\rho}_\mu$ with time
is given by four effects:
\begin{itemize}

\item First by the tendency of the pedestrians to reach
their intended velocity $\vec{v}_\mu^0$ \cite{Alberti,Helbing1}. 
This causes $\hat{\rho}_\mu$
to approach 
\begin{equation}
 \hat{\rho}_\mu^0(\vec{x},\vec{v}_\mu,\vec{v}_\mu^0,t)
:= \delta(\vec{v}_\mu - \vec{v}_\mu^0) \rho_\mu^0(
\vec{x},\vec{v}_\mu^0,t)
\label{iinntt}
\end{equation}
(the equilibrium density in the absence of disturbances)
with a relaxation time 
\begin{displaymath}
 \tau_\mu \equiv \frac{m_\mu}{\gamma_\mu}
\end{displaymath}
(see \cite{Helbing1}). $\rho_\mu^0$ is the density of pedestrians
with intended velocity $\vec{v}_\mu^0$ but arbitrary actual velocity
$\vec{v}_\mu$. $\delta(.)$ denotes the {\sc Dirac} delta function,
which is different from zero only where its argument $(.)$ vanishes.

\item Second by the interaction of pedestrians,
which can be modelled by a 
{\sc Boltzmann}-like {\sl stosszahlansatz} \cite{Helbing2,Kontin2,Kontin1}. 
If we take into account that
the interactions are of short range (in comparison with $r$, see
(\ref{err})), we have:
\begin{eqnarray}
\hat{S}_{\mu \nu} 
&=& \int\!\!\int\!\!\int \hat{\sigma}_{\mu\nu}
(\vec{u}_\mu^*,\vec{u}_\nu^*;\vec{u}_\mu,\vec{u}_\nu;\vec{x},t)
\hat{\rho}_\mu(\vec{x},\vec{u}_\mu^*,t)
\hat{\rho}_\nu(\vec{x},\vec{u}_\nu^*,t)
d^4\vec{u}_\nu \,
d^4\vec{u}_\mu^* \,
d^4\vec{u}_\nu^* \nonumber\\
&-&\int\!\!\int\!\!\int \hat{\sigma}_{\mu \nu}
(\vec{u}_\mu,\vec{u}_\nu;\vec{u}_\mu^*,\vec{u}_\nu^*;\vec{x},t)
\hat{\rho}_\mu(\vec{x},\vec{u}_\mu,t)
\hat{\rho}_\nu(\vec{x},\vec{u}_\nu,t)
d^4\vec{u}_\nu \,
d^4\vec{u}_\mu^* \,
d^4\vec{u}_\nu^* \, . 
\label{stern1}
\end{eqnarray}
This term describes {\sl pair interactions} between pedestrians of types
$\mu$ and $\nu$ occuring with a total rate 
proportional to the densities 
$\hat{\rho}_\mu$ and $\hat{\rho}_\nu$ of both interacting types of motion.
The {\sl relative rate} for pedestrians of types 
$\mu$ and $\nu$ to change their
states from $(\vec{x},\vec{u}_\mu)$,
$(\vec{x},\vec{u}_\nu)$ to 
$(\vec{x},\vec{u}_\mu^*)$, $(\vec{x},\vec{u}_\nu^*)$ due to interactions is
given by $\hat{\sigma}_{\mu \nu}
(\vec{u}_\mu,\vec{u}_\nu;\vec{u}_\mu^*,\vec{u}_\nu^*;\vec{x},t)$.
Assuming that only the actual velocities $\vec{v}_\mu$, $\vec{v}_\nu$ 
and not the intended velocities $\vec{v}_\mu^0$, $\vec{v}_\nu^0$ 
are affected by interactions, we obtain
\begin{eqnarray}
 \hat{\sigma}_{\mu\nu} (\vec{u}_\mu^1,\vec{u}_\nu^1;
\vec{u}_\mu^2, \vec{u}_\nu^2; \vec{x},t) 
= \sigma_{\mu \nu}(\vec{v}_\mu^1,\vec{v}_\nu^1;\vec{v}_\mu^2,\vec{v}_\nu^2)
\delta(\vec{v}_\mu^{0,2}-\vec{v}_\mu^{0,1})\delta(\vec{v}_\nu^{0,2}
- \vec{v}_\nu^{0,1}) 
. \nonumber
\end{eqnarray}
This results in
\begin{eqnarray}
S_{\mu\nu}(\vec{x},\vec{v}_\mu,t)
&:=&\int \hat{S}_{\mu \nu}(\vec{x},\vec{v}_\mu,\vec{v}_\mu^0,t)
d^2 \vec{v}_\mu^0  \nonumber \\ 
&=&
\int\!\!\int \sigma_{\mu\nu}
(\vec{v}_\mu^*,\vec{v}_\nu^*;\vec{v}_\mu,\vec{v}_\nu)
\rho_\mu(\vec{x},\vec{v}_\mu^*,t)
\rho_\nu(\vec{x},\vec{v}_\nu^*,t)
d^2\vec{v}_\nu \,
d^2\vec{v}_\mu^* \,
d^2\vec{v}_\nu^* \nonumber\\
&-&\int\!\!\int \sigma_{\mu \nu}
(\vec{v}_\mu,\vec{v}_\nu;\vec{v}_\mu^*,\vec{v}_\nu^*)
\rho_\mu(\vec{x},\vec{v}_\mu,t)
\rho_\nu(\vec{x},\vec{v}_\nu,t)
d^2\vec{v}_\nu \,
d^2\vec{v}_\mu^* \,
d^2\vec{v}_\nu^* \label{stern2} \\
&=&
\int\!\!\int \sigma_{\mu\nu}^*
(\vec{v}_\mu^*,\vec{v}_\nu^*;\vec{v}_\mu)
\rho_\mu(\vec{x},\vec{v}_\mu^*,t)
\rho_\nu(\vec{x},\vec{v}_\nu^*,t)
d^2\vec{v}_\mu^* \,
d^2\vec{v}_\nu^* \nonumber\\
&-&\int\!\!\int \sigma_{\mu \nu}^*
(\vec{v}_\mu,\vec{v}_\nu;\vec{v}_\mu^*)
\rho_\mu(\vec{x},\vec{v}_\mu,t)
\rho_\nu(\vec{x},\vec{v}_\nu,t)
d^2\vec{v}_\nu \,
d^2\vec{v}_\mu^*  \label{scat1}
\end{eqnarray}
with
\begin{equation}
 \sigma_{\mu\nu}^*(.,.;.)  := \int \sigma_{\mu\nu}
(.,.;.,\vec{v}) d^2\vec{v} \, , \label{scat2}
\end{equation}
which is used later on. (\ref{stern2}) is similar to (\ref{stern1}),
and can be interpreted analogously.
The explicit form of $\sigma_{\mu\nu}^*$ will be
based on a {\sl microscopic model} for the interactions and is discussed
in section 6.

\item Third by pedestrians changing their type $\mu$ of motion 
to another type $\nu$ of motion, for example when turning to the right or left
at a crossing or when turning back (change of the intended direction).
This can be modelled by 
\begin{eqnarray}
\hat{C}_{\mu\nu}(\vec{x},\vec{u}_\mu,t) &=&
\int \hat{\sigma}_\mu^{\nu \mu}(\vec{u}_\nu;\vec{u}_\mu;\vec{x},t)
\hat{\rho}_\nu(\vec{x},\vec{u}_\nu,t)d^4\vec{u}_\nu \nonumber \\
&-& \int \hat{\sigma}_\mu^{\mu \nu}(\vec{u}_\mu;\vec{u}_\nu;\vec{x},t)
\hat{\rho}_\mu(\vec{x},\vec{u}_\mu,t)d^4\vec{u}_\nu  \nonumber
\end{eqnarray}
with a transition rate
proportional to the density of the changing type of motion.
\par
If we assume that for the moment of changing 
the {\sl intended} velocity $\vec{v}_\mu^0$ the 
{\sl actual} velocity $\vec{v}_\mu$ still remains the
same (but of course not thereafter), we have
\begin{displaymath}
 \hat{\sigma}_\mu^{1,2}(\vec{u}_1;\vec{u}_2;\vec{x},t) 
= \hat{\sigma}_\mu^{1,2}(\vec{v}_1^0;\vec{v}_2^0;\vec{x},t)
\delta(\vec{v}_2 -\vec{v}_1) \, .
\end{displaymath}
This results in
\begin{equation}
C_{\mu\nu} := \int \hat{C}_{\mu\nu}(\vec{x},\vec{u}_\mu,t) d^2 \vec{v}_\mu^0
= \sigma_\mu^{\nu \mu}(\vec{x},\vec{v}_\mu,t)
\rho_\nu(\vec{x},\vec{v}_\mu,t) 
- \sigma_\mu^{\mu \nu}(\vec{x},\vec{v}_\mu,t)
\rho_\mu(\vec{x},\vec{v}_\mu,t)  \, ,
\label{change2}
\end{equation}
where
\begin{displaymath}
 \sigma_\mu^{1,2}(\vec{x},\vec{v}_\mu,t) 
:= \int \!\! \int \hat{\sigma}_\mu^{1,2}
(\vec{v}_1^0;\vec{v}_2^0;\vec{x},t) \frac{\hat{\rho}_1
(\vec{x},\vec{v}_\mu,\vec{v}_1^0,t)}{\rho_1(\vec{x},\vec{v}_\mu,t)}
d^2 \vec{v}_1^0 \, d^2 \vec{v}_2^0
\end{displaymath}
and
\begin{equation}
 \rho_\mu(\vec{x},\vec{v}_\mu,t) 
:= \int \hat{\rho}_\mu(\vec{x},\vec{v}_\mu,\vec{v}_\mu^0,t)
d^2 \vec{v}_\mu^0 \, .
\label{rrhhoo}
\end{equation}

\item Fourth by the density gain
$\hat{q}_\mu^+(\vec{x},\vec{v}_\mu,
\vec{v}_\mu^0,t)$ or density loss $\hat{q}_\mu^-(\vec{x},
\vec{v}_\mu,\vec{v}_\mu^0,t)$ per time unit, caused by
pedestrians which enter or leave the system $\cal M$ at a marginal place 
$\vec{x}\in \partial {\cal M}$ (for example a house)
with the intended velocity $\vec{v}_\mu^0 := \vec{v}^0 \in {\cal N}_\mu$ and
the actual velocity $\vec{v}_\mu$:
\begin{equation}
 \hat{q}_\mu(\vec{x},\vec{v}_\mu,
\vec{v}_\mu^0,t) := \hat{q}_\mu^+(\vec{x},\vec{v}_\mu,
\vec{v}_\mu^0,t) - \hat{q}_\mu^-(\vec{x},\vec{v}_\mu,
\vec{v}_\mu^0,t) \, .
\label{qquu}
\end{equation}

\end{itemize}

\section{Macroscopic equations} \label{mac}

For further discussion we need the notations
\begin{eqnarray}
\langle \rho_\mu \rangle 
&:=& \int \rho_\mu(\vec{x},\vec{v}_\mu,t) d^2\vec{v}_\mu 
= \int \hat{\rho}_\mu(\vec{x},\vec{v}_\mu,\vec{v}_\mu^0,t) d^2\vec{v}_\mu 
\, d^2 \vec{v}_\mu^0
\, , \nonumber \\
\langle \psi_\mu(\vec{v}_\mu,\vec{v}_\mu^0)\rangle
&:=& \int \psi(\vec{v}_\mu,\vec{v}_\mu^0)
\frac{\hat{\rho}(\vec{x},\vec{v}_\mu,\vec{v}_\mu^0,t)}
{\langle \rho_\mu \rangle}
d^2\vec{v}_\mu \, d^2 \vec{v}_\mu^0
\, ,  \nonumber \\
\delta \vec{v}_\mu &:=& \vec{v}_\mu 
- \langle \vec{v}_\mu \rangle
\, , \nonumber \\
\delta \vec{v}_\mu^0 &:=& \vec{v}_\mu^0 
- \langle \vec{v}_\mu^0 \rangle
\, , \nonumber \\
\varrho_\mu(\vec{x},\vec{v}_\mu,t) &:=& m_\mu \rho_\mu(\vec{x},\vec{v}_\mu,t)
\, , \label{varrho} \\
p_{\mu, \alpha \beta}
& := & \langle \varrho_\mu \rangle
\langle \delta v_{\mu,\alpha} \delta v_{\mu,\beta} \rangle 
=  \int \delta v_{\mu,\alpha} \delta v_{\mu,\beta} 
\varrho_\mu (\vec{x},\vec{v}_\mu,t) d^2\vec{v}_\mu 
\, , \label{pressure} \\
\vec{j}_{\mu,i} & := & 
\langle \varrho_\mu \rangle \left\langle
\delta \vec{v}_\mu \frac{(\delta v_{\mu,i})^2}{2} \right\rangle 
= \int \delta \vec{v}_\mu \frac{(\delta v_{\mu,i})^2}{2}
\varrho_\mu(\vec{x},\vec{v}_\mu,t) d^2\vec{v}_\mu 
\, , \label{heatflow} \\
\chi_{\mu \nu}\left(\psi_\mu(\vec{v})\right) 
&:=&  \int \!\!\! \int \!\!\! \int \psi_\mu(\vec{v})
\sigma_{\mu \nu}^*(\vec{v}_\mu,\vec{v}_\nu;\vec{v}_\mu^*)
\frac{\rho_\mu(\vec{x},\vec{v}_\mu,t)}{\langle \rho_\mu \rangle}
\frac{\rho_\nu(\vec{x},\vec{v}_\nu,t)}{\langle \rho_\nu \rangle}
d^2\vec{v}_\mu\, d^2\vec{v}_\nu\, d^2\vec{v}_\mu^* \, ,\qquad \label{cchhii}
\\
\chi_{\mu}^{1,2}\left( \psi_\mu (\vec{v}_1)\right) 
&:=& \int \psi_\mu(\vec{v}_\mu) \sigma_\mu^{1,2}(\vec{x},\vec{v}_\mu,t)
\frac{\rho_1(\vec{x},\vec{v}_\mu,t)}
{\langle \rho_1 \rangle}
d^2 \vec{v}_\mu \nonumber \\
&=& \int \psi_\mu(\vec{v}_1) \sigma_\mu^{1,2}(\vec{x},\vec{v}_1,t)
\frac{\rho_1(\vec{x},\vec{v}_1,t)}
{\langle \rho_1 \rangle}
d^2 \vec{v}_1 \, , \nonumber \\
q_\mu(\vec{x},\vec{v}_\mu,t)
& := & \int  \hat{q}_\mu
(\vec{x},\vec{v}_\mu,\vec{v}_\mu^0,t) d^2 \vec{v}_\mu \, d^2 \vec{v}_\mu^0 \, ,
\label{qu} \\
Q_\mu\left( \frac{\psi_\mu(\vec{v}_\mu)}{m_\mu} \right)
&:=& \int \frac{\psi_\mu(\vec{v}_\mu)}{m_\mu} m_\mu
q_\mu(\vec{x},\vec{v}_\mu,t) d^2 \vec{v}_\mu  \, . \nonumber
\end{eqnarray}
Here, $\psi_\mu(\vec{v}_\mu,\vec{v}_\mu^0)$ 
is some arbitrary function of $\vec{v}_\mu$ and $\vec{v}_\mu^0$.
\par
As far as pedestrians of type $\mu$ are concerned,
we are mostly interested in their density $\langle \rho_\mu \rangle$,
their mean velocity
$\langle \vec{v}_\mu \rangle$ and the variance $\langle (\delta v_{\mu,i})^2
\rangle$ of their velocity components $v_{\mu,i}$ 
(at a given place $\vec{x}$ and time $t$). Since it is formally equivalent
and more comparable to fluid dynamics, we shall instead search equations for
the {\sl mass density}
\begin{displaymath}
 \langle \varrho_\mu \rangle := m_\mu \langle \rho_\mu \rangle \, ,
\end{displaymath}
the mean {\sl momentum density}
\begin{displaymath}
 \langle \rho_\mu \rangle \langle m_\mu \vec{v}_\mu \rangle
= \langle \varrho_\mu \rangle \langle \vec{v}_\mu \rangle
\end{displaymath}
and the mean {\sl energy density} (in direction $i$)
\begin{displaymath}
 \langle \epsilon_{\mu,i} \rangle 
:= \langle \rho_\mu \rangle \left\langle \frac{m_\mu}{2} v_{\mu,i}^2
\right\rangle 
= \langle \varrho_\mu \rangle \frac{\langle v_{\mu,i} \rangle^2}{2}
+ \langle \varrho_\mu \rangle \left\langle \frac{(\delta v_{\mu,i})^2}{2}
\right\rangle \, .
\end{displaymath} 
By multiplication of (\ref{kont}) with $\psi_\mu(\vec{v}_\mu)
= m_\mu$, $m_\mu \vec{v}_\mu$ or $m_\mu v_{\mu,i}^2/2$ and
integration over $\vec{u}_\mu$ one obtains the following equations (keeping
in mind that the {\sc Gaussian} surface integrals vanish) (see chapter 2.10
of Ref. \cite{Kontin2}):\alpheqn{formel1}
\begin{eqnarray}
\frac{\partial \langle \varrho_\mu \rangle} {\partial t}
&=& - \frac{\partial}{\partial {}{x}_{\mu,\alpha}}
\left( \langle \varrho_\mu \rangle 
\langle {}{v}_{\mu,\alpha} \rangle \right) 
+ Q_\mu(1) \label{hyd1} \label{formel1a} \\ 
&+& \sum_\nu \left[ \frac{m_\mu}{m_\nu}\langle \varrho_\nu \rangle 
\chi_{\mu}^{\nu \mu}(1)
- \langle \varrho_\mu \rangle \chi_{\mu}^{\mu \nu}(1) \right] \qquad 
\label{typ1}\label{formel1b}
\end{eqnarray}\reseteqn
for the mass density,\alpheqn{formel2} 
\begin{eqnarray}
\frac {\partial 
\left( \langle \varrho_\mu \rangle \langle {}{v}_{\mu,\beta} 
\rangle \right)}{\partial t} 
&=& - \frac{\partial}{\partial {}{x}_{\mu,\alpha}}
\left( \langle \varrho_\mu \rangle 
\langle {}{v}_{\mu,\alpha} \rangle \langle {}{v}_{\mu,\beta}\rangle 
+ p_{\mu,\alpha \beta} \right)
+ \langle \varrho_\mu \rangle  
\frac{{}{f}_{\mu,\beta}}{m_\mu} + Q_\mu(v_{\mu,\beta}) \qquad
\label{hyd2}\label{formel2a} \\
&+& \langle \varrho_\mu \rangle \frac{1}{\tau_\mu} 
\left(\langle{}{v}_{\mu,\beta}^0\rangle
- \langle {}{v}_{\mu,\beta} 
\rangle \right) \label{int2}\label{formel2b} \\
&+& \sum_\nu \langle \varrho_\mu \rangle \langle \varrho_\nu \rangle\frac{1}{m_\nu}
\left[ \chi_{\mu \nu}({}{v}_{\mu,\beta}^*) 
- \chi_{\mu \nu}({}{v}_{\mu,\beta}) \right]  \label{coll2}\label{formel2c} \\ 
&+& \sum_\nu \left[ \frac{m_\mu}{m_\nu}\langle \varrho_\nu \rangle 
\chi_{\mu}^{\nu \mu}\left( v_{\nu,\beta}\right) 
- \langle \varrho_\mu \rangle \chi_{\mu}^{\mu \nu} \left( v_{\mu,\beta}\right)
\right] \label{typ2}\label{formel2d}
\end{eqnarray}\reseteqn
for the momentum density, and\alpheqn{formel3}
\begin{eqnarray}
\frac{\partial \langle \epsilon_{\mu,i} \rangle}{\partial t}
&=& - \frac{\partial}{\partial {}{x}_{\mu,\alpha}} 
\left( \langle {}{v}_{\mu,\alpha} \rangle
\langle \epsilon_{\mu,i} \rangle
+ p_{\mu,\alpha i} \langle {}{v}_{\mu,i} \rangle 
+ {}{j}_{\mu,\alpha, i} \right)
+ \langle \varrho_\mu \rangle \langle {}{v}_{\mu,i} \rangle
\frac{{}{f}_{\mu,i}}{m_\mu} 
+ Q_\mu\left( \frac{v_{\mu,i}^2}{2} \right) \qquad
\label{hyd3}\label{formel3a} \\
&+& \langle \varrho_\mu \rangle 
\frac{1}{\tau_\mu}\left(\langle{}{v}_{\mu,i}^0\rangle^2
- \langle {}{v}_{\mu,i} \rangle^2 \right) \label{int3}\label{formel3b} \\
&+& \langle \varrho_\mu \rangle \frac{1}{\tau_\mu}
\left(\langle (\delta v_{\mu,i}^0)^2 \rangle 
-\langle (\delta v_{\mu,i})^2 \rangle \right) \label{Dance}\label{formel3c} \\
&+& \sum_\nu \langle \varrho_\mu \rangle 
\langle \varrho_\nu \rangle \frac{1}{m_\nu}
\left[ \chi_{\mu \nu}\left(\frac{v_{\mu,i}^{*\mbox{ }2}}{2}\right) 
- \chi_{\mu \nu}\left(\frac{v_{\mu,i}^2}{2}\right) \right]  \label{coll3} 
\label{formel3d} \\ 
&+& \sum_\nu \left[ \frac{m_\mu}{m_\nu} \langle \varrho_\nu \rangle 
\chi_{\mu}^{\nu \mu}
\left( \frac{v_{\nu,i}^2}{2}
\right)
- \langle \varrho_\mu \rangle \chi_{\mu}^{\mu \nu} 
\left( \frac{v_{\mu,i}^2}{2}\right)
\right] \label{typ3}\label{formel3e}
\end{eqnarray}\reseteqn
for the energy density.
Here, we have used the {\sc Einstein}ian 
summation convention to sum over terms 
in which the Greek indices $\alpha$, $\beta$ or $\gamma$ occur twice. 

\subsection{Interpretation} \label{Interpretation}

(\ref{formel1a}), (\ref{formel2a}) and (\ref{formel3a})
are the well known hydrodynamic
equations (see chapters 2.4 and 2.10 of Ref. \cite{Kontin2}). 
(\ref{formel2c}), (\ref{formel3d})
describe the effects of interactions
between two individuals of type $\mu$ and $\nu$ (for details see section 
\ref{inteff}). 
These terms do not vanish as they would do, if a
conservation of momentum and energy would be fulfilled in a {\sl strict} 
sense (see chapter 2.10 of Ref. \cite{Kontin2}). However,
since the individuals
try to approach the intended velocity $\vec{v}_\mu^0$,
there is a {\sl tendency} to restore momentum and energy that is described by 
(\ref{formel2b}), (\ref{formel3b}), (\ref{formel3c}).
\par
(\ref{formel1b}), (\ref{formel2d}), (\ref{formel3e})
are additional terms
due to individuals who change their type of motion. In the following we will
assume the special case that these terms 
as well as the terms $Q_\mu(\psi_\mu(\vec{v}_\mu)/m_\mu)$ due to individuals
entering or leaving the system ${\cal M}$
vanish (by compensation of inflow into $\mu$
and outflow from $\mu$). For concrete situations the quantities
$\chi_\mu^{..}$ and $Q_\mu(.)$ have to be obtained empirically.
\par
$p_{\mu,\alpha \beta}$ is, in thermodynamics, termed the tensor of
{\sl pressure}. $p_{\mu,\alpha\beta}n_\beta$
has the meaning of the force which is used by the individuals of type $\mu$
to change their movement when crossing a line of unit length $l$
(or, more exactly, the meaning of the component of this force
in the direction $\vec{n}$ perpendicular to the line).
$\vec{j}_{\mu,i}$
is, in thermodynamics, called the {\sl heat flow}.
For pedestrians it describes
the tendency of the velocity variance
$\langle (\delta v_{\mu,i})^2 \rangle$
to equalize with time (see (\ref{flow})). The variance
\begin{displaymath}
\theta_{\mu,i} := \langle (\delta v_{\mu,i})^2 \rangle \equiv
k_{_B} T_{\mu,i} / m_\mu
\end{displaymath}
is the thermodynamic equivalent to the {\sl absolute temperature}
$T_{\mu,i}$ in direction $i$.
Approximate expressions for $p_{\mu,\alpha\beta}$ and $\vec{j}_{\mu,i}$
are derived in sections 4 and 5. 

\subsection{Problems of small densities} \label{small}

For pedestrian crowds the densities $\hat{\rho}_\mu$ are usually very small.
As a consequence, equation (\ref{kont}) won't be
fulfilled very well and a discrete formulation would be more appealing
(see \cite{Helbing3}). 
However, we could equivalently start from the continuity equation
(\ref{kont2}), which holds better, because the densities $\rho_\mu$
are only moderately small.
The macroscopic equations will be even better fulfilled,
because they are only equations for the mean values 
$\langle \varrho_\mu \rangle$, $\langle \vec{v}_\mu \rangle$,
$\langle \epsilon_{\mu,i} \rangle$ and could be also set up by
plausibility considerations.
\par
In order to have small fluctuations of the variables 
$\langle \varrho_\mu \rangle$, $\langle \vec{v}_\mu \rangle$,
$\langle \epsilon_{\mu,i} \rangle$ with time,
$\hat{\rho}_\mu$ in equation (\ref{density}) has to be averaged over
a finite area $A$ and a finite volume $V$, which should
be taken sufficiently large. 
If $T$ denotes (appart from fluctuations) the time scale for the
temporal change of $\langle \varrho_\mu \rangle$, 
$\langle \vec{v}_\mu \rangle$ and $\langle \epsilon_{\mu,i} \rangle$, 
these variables can be also averaged over time intervalls 
$\Delta t \ll T$:
\begin{displaymath}
 \overline{\langle \varrho_\mu(\vec{x},t) 
\rangle\langle \psi_\mu(\vec{x},t) \rangle}
:= \frac{1}{\Delta t} \int\limits_{t-\Delta t/2}^{t+\Delta t/2}
\langle \varrho_\mu(\vec{x},t') \rangle 
\langle \psi_\mu(\vec{x},t') \rangle \, dt' \, .
\end{displaymath}
Then, (\ref{formel1}) to (\ref{formel3})
will be proper approximations for the
movement of pedestrians.
\par
Another complication by low densities is, that {\sc Knudsen} {\sl corrections}
have to be taken into account \cite{Henderson}: According to these
corrections, the ``temperature'' $\theta_{\mu,i}$ and the tangential velocity
$\langle v_{\mu,\parallel} \rangle$
change discontinuously at a boundary $\partial {\cal M}$, 
which therefore 
seems to be shifted by a small distance $\xi$ that is
comparable to the mean interaction free path (see \S 14 of Ref. 
\cite{Kontin1}).

\section{Pedestrians in equilibrium} \label{pedinequ}

In order to calculate $p_{\mu,\alpha\beta}$, $\vec{j}_{\mu,i}$ and
$\chi_{\mu\nu}$ we need the explicit form of
$\rho_\mu$ (see (\ref{rrhhoo}), and (\ref{varrho}) to
(\ref{cchhii})). $\rho_\mu$ is the density of individuals of type $\mu$ at
place $\vec{x}$ and time $t$ having the actual velocity $\vec{v}_\mu$ but
arbitrary intended velocity $\vec{v}_\mu^0$ (see (\ref{rrhhoo})). It is 
directly measurable in pedestrian crowds.
By integration of (\ref{kont})
over $\vec{v}_\mu^0$ we obtain the theoretical dependence
\begin{eqnarray}
 \frac{d\rho_\mu}{dt} &\equiv &
\frac{\partial \rho_\mu}{\partial t} 
+ \nabla_{\vec{x}} \, (\rho_\mu \vec{v}_\mu) 
+ \nabla_{\vec{v}_\mu} \left(\rho_\mu 
\frac{\vec{f}_\mu}{m_\mu}\right)
\nonumber \\
&=& 
\frac{\rho_\mu^0-\rho_\mu}{\tau_\mu} 
+ \sum_\nu S_{\mu \nu} \label{kont2} \\
&+& \sum_\nu C_{\mu \nu}
+ q_\mu \nonumber 
\end{eqnarray}
(compare to \cite{Prig,Pav}).
So, the temporal development of the density $\rho_\mu$
is given by a tendency to walk with the intended velocity $\vec{v}_\mu^0$
(see (\ref{iinntt})), by the effects $S_{\mu\nu}$ of pair interactions 
(see (\ref{scat1})), by the effects $C_{\mu\nu}$ of pedestrians
changing their type of motion (see (\ref{change2})) 
and by the effect $q_\mu$ of pedestrians
entering or leaving the system $\cal M$ (see (\ref{qquu}), (\ref{qu})).
The last two effects shall be be neglected in the following
(see the comment in section \ref{Interpretation}).
\par
Equation (\ref{kont2}) can be solved in a suitable approximation by
the recursive method of {\sc Chapman} and {\sc Enskog} 
\cite{Chapman,Enskog}. The lowest
order approximation presupposes the condition $d\rho_\mu^e/dt = 0$ 
of {\sl local equilibrium}, which is approximately fulfilled by the 
{\sc Gauss}ian distribution
\begin{eqnarray}
 \rho_\mu^e(\vec{x},\vec{v}_\mu,t) 
&=& \langle \rho_\mu \rangle \cdot \frac{1}{2\pi b_\mu \theta_{\mu,\parallel}}
\mbox{e}^{-[(v_{\mu,\parallel} 
- \langle v_{\mu,\parallel} \rangle
)^2
/ (2\theta_{\mu,\parallel})
+ (v_{\mu,\perp} 
- \langle v_{\mu,\perp} \rangle
)^2/(2\theta_{\mu,\perp})]}
\label{equi}
\end{eqnarray}
according to empirical data \cite{Rennen,Frauen,Soldaten}. 
\begin{displaymath}
 (b_\mu)^2 := \frac{\theta_{\mu,\perp}}{\theta_{\mu,\parallel}} \le 1
\end{displaymath}
describes the fact that the velocity variance $\theta_{\mu,i}$
perpendicular ($\perp$) to the
mean intended direction of movement $\langle \vec{v}_\mu^0 \rangle$ is
normally less than parallel ($\parallel$) to it \cite{Soldaten}.
\par
For each type $\mu$ of motion let us perform a particular transformation
\begin{displaymath}
 \vec{x} \longrightarrow  
\vec{X}_\mu :=\left(
\begin{array}{c}
x_{\mu,\parallel} \\
x_{\mu,\perp}/b_\mu
\end{array} \right)\, , 
\end{displaymath}
\begin{displaymath} 
\vec{v}_\mu
\longrightarrow 
\vec{V}_\mu := \left(
\begin{array}{c}
v_{\mu,\parallel} \\
v_{\mu,\perp}/b_\mu
\end{array} \right) \, , 
\end{displaymath}
\begin{displaymath} 
\vec{f}_\mu
\longrightarrow 
\vec{F}_\mu := \left(
\begin{array}{c}
f_{\mu,\parallel} \\
f_{\mu,\perp}/b_\mu
\end{array} \right) \, , 
\end{displaymath}
\begin{displaymath}
\langle \varrho_{\mu,\perp}(\vec{x}) \rangle 
:= \frac{1}{\Delta x_\parallel} \int \limits_{\Delta x_\parallel}
\langle \varrho_\mu \rangle d x_{\mu,\parallel} \longrightarrow 
b_\mu \langle \varrho_{\mu,\perp}(\vec{X}_\mu)
\rangle \, , 
\end{displaymath}
\begin{displaymath}
p_{\mu,\alpha \beta} =: \epsilon_{\mu,\alpha\gamma} P_{\mu,\gamma \beta}
\longrightarrow  P_{\mu,\alpha \beta} \, ,
\end{displaymath}
\begin{displaymath}
j_{\mu,\alpha,i}=:\epsilon_{\mu,\alpha\beta}J_{\mu,\beta,i} 
\longrightarrow  J_{\mu,\alpha,i}
\end{displaymath}
with 
\begin{displaymath}
 \underline{\epsilon}_\mu \equiv (\epsilon_{\mu,\alpha \beta}) := \left(
\begin{array}{cc}
1 & 0 \\
0 & b_\mu^2
\end{array} \right) \, .
\end{displaymath}
This transformation stretches the direction perpendicular to $\langle
\vec{v}_\mu^0 \rangle$ by the factor $1/b_\mu$ and
simplifies equations (\ref{formel1}) to (\ref{formel3})
to {\sl isotropic} ones (that means 
to equations with local rotational symmetry).
With
\begin{displaymath}
 \theta_\mu := \theta_{\mu,\parallel}
\end{displaymath}
we get 
\begin{equation}
P_{\mu,\alpha\beta}^e = \langle \varrho_\mu \rangle \theta_\mu 
\delta_{\alpha \beta} =: P_\mu^e \delta_{\alpha\beta} \, , \label{ideal}
\end{equation}
for the pressure and
\begin{equation}
J_{\mu,\alpha,i}^e = 0  \label{Strom}
\end{equation}
for the ``heat flow'' (see chapter 2.10 of Ref. \cite{Kontin2}). 
In addition, the {\sc Euler} equations 
\begin{eqnarray}
\frac{d \langle \varrho_\mu \rangle}{d t} &:=&
\frac{\partial \langle \varrho_\mu \rangle}{\partial t}
+ \langle V_{\mu,\alpha} \rangle \frac{\partial \langle \varrho_\mu \rangle}{\partial X_{\mu,\alpha}}
= 
- \langle \varrho_\mu \rangle \frac{\partial \langle V_{\mu,\alpha} \rangle}{\partial X_{\mu,\alpha}}
\, ,\label{euler1} \\
\frac{d \langle V_{\mu,\beta} \rangle}{d t} &:=&
\frac{\partial \langle V_{\mu,\beta} \rangle}{\partial t}
+ \langle V_{\mu,\alpha} \rangle \frac{\partial \langle V_{\mu,\beta}\rangle}
{\partial X_{\mu,\alpha}} =
- \frac{1}{\langle \varrho_\mu \rangle}
\frac{\partial P_{\mu,\alpha \beta}^e}{\partial X_{\mu,\alpha}}
+ \frac{F_{\mu,\beta}}{m_\mu}
\, ,\label{euler2} \\
\frac{d \theta_{\mu}}{d t} &:=&
\frac{\partial \theta_{\mu}}{\partial t} 
+\langle V_{\mu,\alpha} \rangle \frac{\partial \theta_{\mu}}{\partial X_{\mu,\alpha}}
=- \theta_{\mu} \frac{\partial \langle V_{\mu,\alpha} \rangle}
{\partial X_{\mu,\alpha}} 
\label{euler3}
\end{eqnarray}
can be derived from (\ref{kont}) (see chapter 16.2 of Ref. \cite{Stumpf}).

\subsection{Behavior on a dance floor}

\begin{figure}[htbp]
\unitlength1cm
\begin{center}
\begin{picture}(6,6)
\put(0.9,0.8){\circle{0.43}}
\put(0.4,1.1){\circle{0.43}}
\put(1.5,0.5){\circle{0.43}}
\put(0.5,0.4){\circle{0.43}}
\put(1.1,0.1){\circle{0.43}}
\put(1.9,0.1){\circle{0.43}}
\put(2.4,-0.1){\circle{0.43}}
\put(2.8,0.3){\circle{0.43}}
\put(3.3,0){\circle{0.43}}
\put(3.8,-0.2){\circle{0.43}}
\put(4.1,0.3){\circle{0.43}}
\put(4.5,-0.1){\circle{0.43}}
\put(4.8,0.4){\circle{0.43}}
\put(5.4,0.3){\circle{0.43}}
\put(5.1,0.8){\circle{0.43}}
\put(5.5,1.2){\circle{0.43}}
\put(5.8,0.8){\circle{0.43}}
\put(6,1.3){\circle{0.43}}
\put(5.7,1.8){\circle{0.43}}
\put(6,2.3){\circle{0.43}}
\put(6.3,2.8){\circle{0.43}}
\put(5.8,3.2){\circle{0.43}}
\put(6.1,3.6){\circle{0.43}}
\put(5.7,4.1){\circle{0.43}}
\put(6.3,4.1){\circle{0.43}}
\put(6,4.5){\circle{0.43}}
\put(6.1,5){\circle{0.43}}
\put(5.7,5.4){\circle{0.43}}
\put(5.4,4.9){\circle{0.43}}
\put(4.7,5.6){\circle{0.43}}
\put(5.2,5.7){\circle{0.43}}
\put(4.6,6.1){\circle{0.43}}
\put(4.1,6){\circle{0.43}}
\put(3.7,5.7){\circle{0.43}}
\put(3.5,6.3){\circle{0.43}}
\put(3,5.9){\circle{0.43}}
\put(3,6.5){\circle{0.43}}
\put(2.5,6.2){\circle{0.43}}
\put(2.1,5.8){\circle{0.43}}
\put(1.7,6.1){\circle{0.43}}
\put(1.1,5.9){\circle{0.43}}
\put(1.1,5.3){\circle{0.43}}
\put(0.5,5.9){\circle{0.43}}
%
\put(0.6,5.3){\circle{0.43}}
\put(0,5.2){\circle{0.43}}
\put(0.3,4.7){\circle{0.43}}
\put(-0.1,4.3){\circle{0.43}}
\put(0.2,3.9){\circle{0.43}}
\put(0,3.4){\circle{0.43}}
\put(-0.3,3){\circle{0.43}}
\put(0.1,2.5){\circle{0.43}}
\put(-0.3,2.1){\circle{0.43}}
\put(0.3,1.8){\circle{0.43}}
\put(-0.2,1.4){\circle{0.43}}
\put(0,0.8){\circle{0.43}}
%
%
\put(1.5,1.8){\circle*{0.43}}
\put(1.5,1.8){\vector(-1,3){0.2}}
\put(1.5,1.8){\vector(1,-3){0.2}}
\put(2.7,1){\circle*{0.43}}
\put(2.7,1){\vector(1,1){0.43}}
\put(2.7,1){\vector(-1,-1){0.43}}
\put(4,2){\circle*{0.43}}
\put(4,2){\vector(-1,1){0.43}}
\put(4,2){\vector(1,-1){0.43}}
\put(3,3){\circle*{0.43}}
\put(3,3){\vector(2,1){0.6}}
\put(3,3){\vector(-2,-1){0.6}}
\put(2,4.5){\circle*{0.43}}
\put(2,4.5){\vector(-1,2){0.3}}
\put(2,4.5){\vector(1,-2){0.3}}
\put(3.2,4.8){\circle*{0.43}}
\put(3.2,4.8){\vector(1,1){0.43}}
\put(3.2,4.8){\vector(-1,-1){0.43}}
\put(4,4.1){\circle*{0.43}}
\put(4,4.1){\vector(1,0){0.7}}
\put(4,4.1){\vector(-1,0){0.7}}
\put(5.1,3.2){\circle*{0.43}}
\put(5.1,3.2){\vector(0,1){0.7}}
\put(5.1,3.2){\vector(0,-1){0.7}}
\put(1,3.8){\circle*{0.43}}
\put(1,3.8){\vector(1,2){0.3}}
\put(1,3.8){\vector(-1,-2){0.3}}
\end{picture}
\end{center}
\caption[]{Behavior on a dance floor: Dancing individuals 
(filled circles) show a lower density than the individuals
standing around (empty circles), since they intend to move
with a greater velocity variance.\label{dance}}
\end{figure}
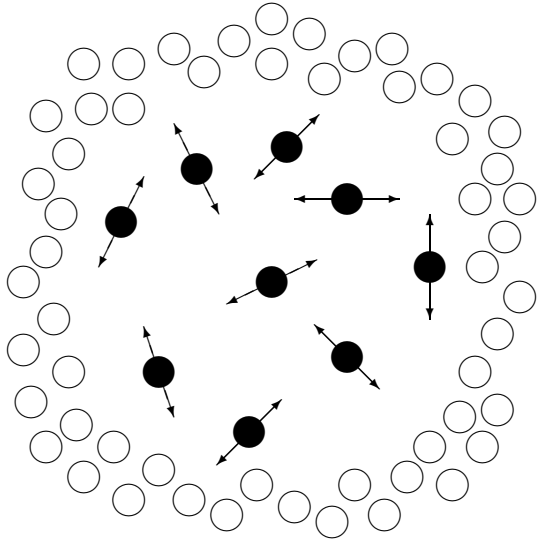

On a dance floor like that of a discotheque, two types of motion can be found:
Type 1 describing dancing individuals, type 2 describing
individuals standing around and
looking on. We can assume to have an isotropic case, that means $b_\mu = 1$
and $\theta_{\mu,\parallel} = \theta_{\mu,\perp}$.
According to (\ref{formel3c}), the variance 
$\langle (\delta v_{\mu,i}^0)^2 \rangle$
of the {\sl intended} velocities $v_{\mu,i}^0$
is {\sl causal} for the temperature $\theta_{\mu,i}$,
that means for the variance $\langle (\delta v_{\mu,i})^2 \rangle$
of the actual velocities $v_{\mu,i}$. So for the
temperatures $\theta_1$, $\theta_2$ of individuals dancing and individuals
standing around it holds
\begin{displaymath}
 \theta_1 > \theta_2 \, ,
\end{displaymath}
since the dancing individuals intend to move with higher variance
\mbox{$\langle (\delta v_{1,i}^0)^2 \rangle >
\langle (\delta v_{2,i}^0)^2 \rangle$}. (This is even so in the case of
equilibrium, because we have to take
the effect of the {\sc Knudsen} corrections into account,
see section \ref{small}.) The equilibrium condition of equal pressure
\begin{displaymath}
 P_1^e = P_2^e
\end{displaymath}
now leads to
\begin{displaymath}
 \langle \varrho_1 \rangle = \frac{\theta_2}{\theta_1} 
 \langle \varrho_2 \rangle < \langle \varrho_2 \rangle 
\end{displaymath}
(see (\ref{ideal})).
Therefore, the dancing individuals will have less density than the individuals
standing around (see figure \ref{dance}). 
This effect can actually be observed.

\subsection{Propagation of density waves}

In a nearly homogenuous pedestrian crowd
with small density variations one can assume
\begin{displaymath}
 \vec{V}_\mu \cdot \nabla_{\vec{X}_\mu} 
 \langle \varrho_\mu \rangle \approx 0 \, , \quad
 \langle \vec{V}_\mu \rangle \nabla_{\vec{X}_\mu}
 \langle \vec{V}_\mu \rangle \approx \vec{0} \, , \quad
 \langle \vec{V}_\mu \rangle \nabla_{\vec{X}_\mu} \theta_\mu \approx 0
 \, , \quad F_{\mu,\beta} \approx 0 \, .
\end{displaymath}
From the {\sc Euler} equations (\ref{euler1}) to (\ref{euler3})
the equation 
\begin{displaymath}
\frac{\partial^2 \langle \varrho_\mu \rangle}{\partial t^2}
- \Delta_{\vec{X}_\mu} P_\mu^e = 0 
\end{displaymath}
can be derived, then (see chapter 16.2 of Ref. \cite{Stumpf}).
Subtracting $\langle \varrho_\mu \rangle / \theta_\mu \times$(\ref{euler3}) 
from (\ref{euler1}) and making use of
(\ref{ideal}) and (\ref{Strom}), the {\sl adiabatic law}
\begin{displaymath}
 \frac{P_\mu^e}{\langle \varrho_\mu \rangle^2} = const.
\end{displaymath}
can be shown, which leads to the {\sl linear 
wave equation}
\begin{equation}
 \langle \varrho_\mu \rangle
\kappa_\mu^S \frac{\partial^2}{\partial t^2}\langle \varrho_\mu \rangle 
-\Delta_{\vec{X}_\mu} \langle \varrho_\mu \rangle =0 
\label{wave}
\end{equation}
with the {\sl adiabatic compressibility}
\begin{displaymath}
\kappa_\mu^S := \frac{1}{\langle \varrho_\mu \rangle}\left(
\frac{\partial P_\mu^e}{\partial \langle \varrho_\mu \rangle} \right)_S
= \frac{1}{2} \frac{\langle \varrho_\mu  \rangle}{P_\mu^e} =
\frac{1}{2\langle \varrho_\mu \rangle \theta_\mu}  
\end{displaymath}
(see chapter 16.2 of Ref. \cite{Stumpf}).
(For the description of nonlinear waves see \cite{Whit2} and
chapter 2.1 of Ref. \cite{Whit1}.)
Equation (\ref{wave}) describes the propagation of density waves with velocity
\begin{equation}
c_\mu = \frac{1}{\sqrt{\langle \varrho_\mu \rangle \kappa_\mu^S}}
\, . \label{veloc}
\end{equation}
On the other hand, the velocity of propagation is given by the mean distance
$d_\mu = 1/\sqrt{\langle \rho_\mu \rangle}$ of the succeeding individual
devided by its mean reaction time $\zeta_\mu$:
\begin{displaymath}
 c_\mu = \frac{1}{\sqrt{\langle \rho_{\mu} 
\rangle} \zeta_\mu} \, .
\end{displaymath}
Inserting this into (\ref{veloc}), it follows that for
small densities $\langle \rho_\mu \rangle$ the 
adiabatic compressibility grows with the mean {\sl reaction time}
$\zeta_\mu$ of individuals according to
\begin{displaymath}
\kappa_\mu^S = \frac{(\zeta_\mu)^2}{m_\mu}  \, .
\end{displaymath}

\section{Nonequilibrium equations}

In cases of deviations
\begin{equation}
\delta \rho_\mu(\vec{X}_\mu,\vec{V}_\mu,t) :=
\rho_\mu(\vec{X}_\mu,\vec{V}_\mu,t) - \rho_\mu^e(\vec{X}_\mu,\vec{V}_\mu,t) 
\label{devia}
\end{equation}
from local equilibrium $\rho_\mu^e(\vec{X}_\mu,\vec{V}_\mu,t)
:= \rho_\mu^e(\langle \rho_\mu \rangle, \langle \vec{V}_\mu \rangle,
\theta_\mu)$, 
we have to find a higher order approximation of equation (\ref{kont2})
than in section \ref{pedinequ}. If the deviations $\delta \rho_\mu$ remain
small compared with $\rho_\mu$, we 
can linearize equation (\ref{kont2})
around $\rho_\mu^e$ and get 
\begin{displaymath}
\frac{d \rho_\mu^e}{dt} \approx
\frac{d \rho_\mu^e}{dt} + \frac{d \delta \rho_\mu}{dt}
= \frac{d\rho_\mu}{dt}
\approx - \frac{\delta \rho_\mu}{\tau_\mu}
+ \sum_\nu \frac{\delta \rho_\nu}{\tau_{\mu\nu}} 
\end{displaymath}
(see chapter 15 of Ref. \cite{Jaeckle}).
Here, $\tau_{\mu\nu}$ is the mean interaction free time between
an individual of type $\mu$ and individuals of type
$\nu$ (see (\ref{interrates}) and chapter 16.2 of Ref. \cite{Stumpf}). 
From (\ref{equi}) and
(\ref{ideal}) to (\ref{euler3}) one finds \cite{Stumpf}
\begin{eqnarray}
 \frac{d \rho_\mu^e}{dt} &=&
\frac{\partial \rho_\mu^e }{\partial \langle \rho_\mu \rangle}
\frac{d\langle \rho_\mu \rangle}{dt}
+ \left(\nabla_{\langle \vec{V}_\mu \rangle} \rho_\mu^e \right) \cdot
\frac{d \langle \vec{V}_\mu \rangle}{dt}
+ \frac{\partial \rho_\mu^e}{\partial \theta_\mu}\frac{d\theta_\mu}{dt}
\nonumber \\
&=&\rho_\mu^e \left[ \frac{\delta \vec{V}_\mu}{\theta_\mu} \cdot
\nabla_{\vec{X}_\mu} \theta_\mu\left( \frac{(\delta \vec{V}_\mu)^2}
{2\theta_\mu} - 2 \right) \right] \nonumber \\
&+& \rho_\mu^e \left[ \frac{1}{\theta_\mu} \left(
\delta V_{\mu,\alpha} \frac{\partial \langle V_{\mu,\beta} \rangle}
{\partial X_{\mu,\alpha}} \delta V_{\mu,\beta} 
-\frac{(\delta \vec{V}_\mu)^2}{2} \nabla_{\vec{X}_\mu}
\langle \vec{V}_\mu \rangle \right)\right] \, . \nonumber
\end{eqnarray}
If 
\begin{displaymath}
(\tau_\mu \delta_{\mu\nu} + \overline{\tau}_{\mu\nu})
\end{displaymath}
denotes the inverse matrix of
\begin{displaymath}
 \left( \frac{1}{\tau_\mu} \delta_{\mu\nu} - \frac{1}{\tau_{\mu\nu}}\right)\, ,
\end{displaymath}
the relation 
\begin{equation}
 \delta \rho_\mu = - \tau_\mu \frac{d\rho_\mu^e}{dt}
- \sum_\nu \overline{\tau}_{\mu\nu} \frac{d\rho_\nu^e}{dt} \label{rela}
\end{equation}
leads, because of $\rho_\mu = \rho_\mu^e + \delta \rho_\mu$, 
to the corrected tensor of pressure
\begin{displaymath}
P_{\mu,\alpha \beta} = P_\mu^e \delta_{\alpha\beta} 
- \eta_\mu \Lambda_{\mu,\alpha\beta} 
- \sum_\nu \eta_{\mu\nu} \Lambda_{\nu,\alpha\beta}
\end{displaymath}
(see (\ref{pressure})) and the corrected heat flow 
\begin{equation}
 J_{\mu,\alpha} 
= - \kappa_\mu \frac{\partial \theta_\mu}{\partial X_{\mu,\alpha}}
 - \sum_\nu \kappa_{\mu\nu}  \frac{\partial \theta_\nu}{\partial X_{\nu,\alpha}}
 \label{flow}
\end{equation}
(see (\ref{heatflow})). Here,
\begin{displaymath}
 \Lambda_{\mu, \alpha \beta}:= \glossary{$\Lambda_{\mu, \alpha\beta}$} \left(
\frac{\partial \langle V_{\mu,\alpha}
\rangle}{\partial X_{\mu,\beta}}
+\frac{\partial \langle V_{\mu,\beta}
\rangle}{\partial X_{\mu,\alpha}}-\frac{\partial \langle V_{\mu,\alpha} \rangle}
{\partial X_{\mu,\alpha}} \delta_{\alpha \beta}\right) 
\end{displaymath}
is the {\sl tensor of strain},
\begin{displaymath}
 \eta_\mu = \tau_\mu \theta_\mu \langle \varrho_\mu \rangle \, , \qquad
 \eta_{\mu\nu} = \overline{\tau}_{\mu\nu} 
\theta_\nu \langle \varrho_\nu \rangle
\end{displaymath}
are coefficients of the {\sl shear viscosity}, and
\begin{displaymath}
\kappa_\mu = 2 \tau_\mu \theta_\mu \langle \varrho_\mu \rangle \, , \qquad
\kappa_{\mu\nu} = 2 \overline{\tau}_{\mu\nu} 
\theta_\nu \langle \varrho_\nu \rangle
\end{displaymath}
are coefficients of the {\sl thermal conductivity}.
\par
Note, that the effect of restoring the local equilibrium distribution
$\rho_\mu^e$ results from the tendency to approach the 
intended velocity distribution 
$\rho_\mu^0(\vec{X}_\mu,\vec{V}_\mu,t)$ with a time constant
$\tau_\mu$, but not from interaction
processes as usual (see chapter 13.3 of Ref. \cite{Reif}). 
Therefore the viscosity $\eta_\mu$
is dependent on the density $\langle \rho_\mu \rangle$ in contrast
to ordinary fluids (see pages 323 and 327 in Ref. \cite{Stumpf}).
For vanishing densities $\langle \rho_\nu \rangle \longrightarrow 0$
the interaction rates $1/\tau_{\mu\nu}$ become 
negligible (see (\ref{interrates}))
and $\overline{\tau}_{\mu\nu}$, $\eta_{\mu\nu}$, $\kappa_{\mu\nu}$
vanish in comparison with $\tau_\mu$, $\eta_\mu$, $\kappa_\mu$, respectively.
According to (\ref{rela}), the deviation $\delta \rho_\mu$ from
the local equilibrium distribution $\rho_\mu^e$ and, therefore,
the viscosity and the thermal conductivity are consequences of
finite relaxation times $\tau_\mu$, $\tau_{\mu\nu}$.

\subsection{Effect of viscosity}

\begin{figure}[htbp]
\unitlength1cm
\begin{picture}(12,8)
\put(0.5,0.5){\vector(0,1){6}}                                                 
\put(0.4,3.5){\line(1,0){0.2}}
\put(0,7.5){$X_{\mu,\perp}$}
\put(0,3.4){0}
\put(1.5,0){\vector(1,0){10.5}}
\put(12.5,-0.1){$\langle V_{\mu,\parallel} \rangle$}
\end{picture}
\caption[]{Effect of viscosity (internal friction): Ordinary fluids show a
parabolic velocity profile (broken line). In contrast to this, a hyperbolic
velocity profile is expected for pedestrian crowds (solid line). Whereas
in ordinary fluids the internal friction is compensated by a pressure gradient,
for pedestrian crowds
this role is played by the accelerating effect of the intended 
velocity. Due to the {\sc Knudsen} corrections 
the fluid slips at the boundary, that means the {\sl effective} width is 
greater than the {\sl actual} width.\label{profile}}
\end{figure}
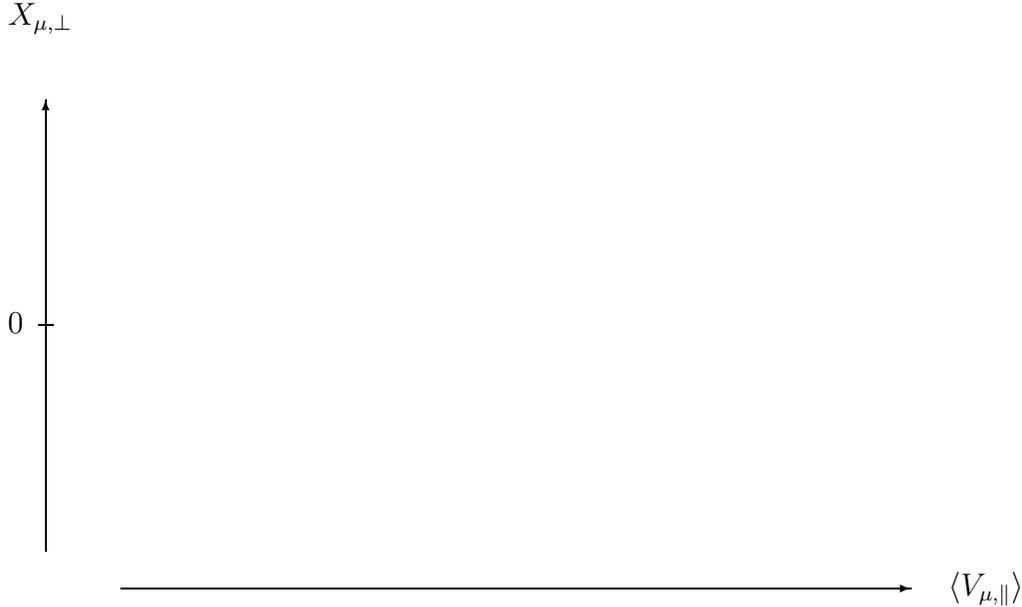

For pedestrians the effect of viscosity is not compensated by a 
gradient of pressure as in ordinary fluids, but instead by the 
tendency of pedestrians to reach their intended velocity
described by (\ref{formel2b}). In the case of 
a stationary flow in one direction (that means of one type of
motion) parallel to the boundaries $\partial {\cal M}$ we have
essentially the equation
\begin{equation}
0 = 
\frac{\partial \langle \varrho_\mu \rangle\langle V_{\mu,\parallel} \rangle}
{\partial t} = \eta_\mu \frac{\partial^2}{\partial X_{\mu,\perp}^2}
\langle V_{\mu,\parallel} \rangle + \langle \varrho_\mu \rangle
\frac{1}{\tau_\mu} \left(\langle V_{\mu,\parallel}^0\rangle 
- \langle V_{\mu,\parallel} \rangle \right) \, ,
\label{frict}
\end{equation}
if $\eta_\mu \gg \eta_{\mu\mu}$ (see (\ref{formel2a}), (\ref{formel2b})).
For a lane of effective width $2W$ 
(with the origin $X_{\mu,\perp}=0$ in the middle)
equation (\ref{frict}) has the {\sl hyperbolic} solution
\begin{equation}
 \langle V_{\mu,\parallel} \rangle = \langle V_{\mu,\parallel}^0 \rangle 
\left[ 1-\frac{\cosh(X_{\mu,\perp}/D_\mu)}
{\cosh(W/D_\mu)} \right]
\label{hyperbolic}
\end{equation}
with a {\sl boundary layer} of width
\begin{displaymath}
D_\mu =\sqrt{\frac{\eta_\mu \tau_\mu}
{\langle\varrho_\mu \rangle}} = \tau_\mu \sqrt{\theta_\mu} \, .
\end{displaymath}
In comparison with this, a pressure gradient
\begin{displaymath}
\frac{\partial P_{\mu,_{\parallel\parallel}}^e}{\partial X_{\mu,\parallel}}
:= \frac{\Delta P_{\mu}^e}{L}
\end{displaymath}
generating the driving force would lead instead to the {\sl parabolic}
solution
\begin{equation}
 \langle V_{\mu,\parallel} \rangle = 
\frac{\Delta P_{\mu}^e}{\eta_\mu L}
(W^2-X_{\mu,\perp}^2) \, ,
\label{parabolic}
\end{equation}
and the mean tangential velocity $\langle V_{\mu,\parallel} \rangle$
would depend on the length $L$ of the lane
(see chapter 3.3 of Ref. \cite{Hagen}). 
The hyperbolic solution
(\ref{hyperbolic}) and the parabolic solution (\ref{parabolic}) are 
depicted in figure \ref{profile}.

\section{Effects of interactions} \label{inteff}

The {\sl scattering rates} $\sigma_{\mu\nu}^*$ of interactions 
(see (\ref{scat1}), (\ref{scat2}))
are proportional to the relative velocity $\| \vec{v}_\mu - \vec{v}_\nu \|$
of the interacting pedestrians
and to the {\sl scattering cross section} $l_{\mu\nu}$ (which is a length
here and of order of a pedestrian's stride) \cite{Kontin2}:
\begin{displaymath}
\sigma_{\mu \nu}^*(\vec{v}_\mu,\vec{v}_\nu;
\vec{v}_\mu^*)
= l_{\mu \nu}\|\vec{v}_\mu-\vec{v}_\nu\| P_{\mu \nu}(\vec{v}_\mu,
\vec{v}_\nu;\vec{v}_\mu^*) \, . 
\end{displaymath}
It is
\begin{eqnarray}
 \frac{1}{\tau_{\mu\nu}} &:=& \frac{1}{\langle \rho_\mu \rangle}
\int\!\!\int \rho_\mu(\vec{x},\vec{v}_\mu,t)\rho_\nu(\vec{x},\vec{v}_\nu,t)
l_{\mu\nu} \|\vec{v}_\mu -\vec{v}_\nu\| d^2\vec{v}_\mu \, d^2\vec{v}_\nu
\nonumber \\
&=& \langle \rho_\nu \rangle l_{\mu\nu} \langle \|\vec{v}_\mu 
- \vec{v}_\nu \| \rangle \label{interrates}
\end{eqnarray}
the mean rate of interactions of an individual of type $\mu$ with 
individuals of type $\nu$, and $\tau_{\mu\nu}$ the corresponding
mean interaction free
time \cite{Kontin2}. For the mean relative velocity $\langle \|
\vec{v}_\mu - \vec{v}_\nu \| \rangle$ (see figure
\ref{meanrel}) the following limits can be calculated 
by making use of (\ref{equi}), (\ref{devia}) and neglecting terms
of order $O(\delta \rho_\mu)$:
\begin{equation}
 \langle \| \vec{v}_\mu - \vec{v}_\nu \| \rangle \approx \left\{
\begin{array}{ll}
\sqrt{\pi \theta_\mu} & \mbox{if } \langle \vec{v}_\mu \rangle
\approx \langle \vec{v}_\nu \rangle,\, \theta_\mu \approx \theta_\nu \\
\| \langle \vec{v}_\mu \rangle - \langle \vec{v}_\nu \rangle \| &
\mbox{if } \| \langle \vec{v}_\mu \rangle - \langle \vec{v}_\nu \rangle \|
\gg \theta_\mu,\theta_\nu \, .
\end{array} \right. \label{jams}
\end{equation}
\begin{figure}[htbp]
\unitlength1cm
\begin{picture}(14,11)
\put(0,10.5){$\langle \| \vec{v}_\mu - \vec{v}_\nu \|\rangle$}
\put(12,0){$\| \langle \vec{v}_\mu \rangle -
\langle \vec{v}_\nu \rangle \|$}
\end{picture}
\caption[]{The mean relative velocity $\langle \| \vec{v}_\mu - \vec{v}_\nu \|
\rangle$ in dependence of $\| \langle \vec{v}_\mu \rangle -
\langle \vec{v}_\nu \rangle \|$ for the special case $\theta_{\mu,i} = 1
= \theta_{\nu,i}$.\label{meanrel}}
\end{figure}
\par
Let us introduce 
\begin{equation}
 \tau_{\mu\nu}^* =
 \tau_{\mu\nu}^*(\langle\vec{v}_\mu \rangle,\langle \vec{v}_\nu \rangle,
\theta_\mu,\theta_\nu) := \tau_{\mu\nu} \langle \varrho_\nu \rangle
= \tau_{\mu\nu} m_\nu \langle \rho_\nu \rangle \, ,
\end{equation}
and the {\sl total rate of interactions} 
\begin{equation}
 \frac{1}{\hat{\tau}_\mu} 
:= \sum_\nu \frac{1}{\tau_{\mu\nu}} \, .  
\label{freetime}
\end{equation}
Then,
\begin{equation}
 r_\mu = \mbox{e}^{-\Delta t_\mu/\hat{\tau}_\mu} \label{rrr}
\end{equation}
is the probability for the possibility to pass 
an individual to the right or left, if this would need
an interaction free time of at least $\Delta t_\mu$ 
(see chapter 12.1 of Ref. \cite{Reif}).
\begin{displaymath}
 P_{\mu\nu} (\vec{v}_\mu,\vec{v}_\nu;
\vec{v}_\mu^*)= \sum_k P_{\mu\nu}^k (\vec{v}_\mu,\vec{v}_\nu;
\vec{v}_\mu^*)
\end{displaymath}
is the probability, that two individuals of types $\mu$ and $\nu$
have velocities $\vec{v}_\mu$ and $\vec{v}_\nu$ before their interaction,
and the individual of type $\mu$ has the velocity $\vec{v}_\mu^*$ 
thereafter. We shall distinguish three types $k$ of interaction: 
\par
If an
individual of type $\mu$ is hindered by another individual of type $\nu$,
it tries to pass the other to the right with probability $p_{\mu\nu}$ or to
the left with probability $1-p_{\mu\nu}$:
\begin{displaymath}
P_{\mu \nu}^1(\vec{v}_\mu,\vec{v}_\nu;\vec{v}_\mu^*)
= r_\mu\left[p_{\mu \nu}\delta(\vec{v}_\mu^*-\underline{S}_{\beta_{\mu\nu}}
\vec{v}_\mu)
+(1-p_{\mu \nu})\delta(\vec{v}_\mu^*-\underline{S}_{\beta_{\mu\nu}}^{-1} 
\vec{v}_\mu) \right] \, . 
\end{displaymath}
$\vec{v}_\mu^*=\underline{S}_{\beta_{\mu\nu}}\vec{v}_\mu$ 
describes a rotation of velocity
$\vec{v}_\mu$ to the right side
by an angle $\beta_{\mu\nu}$ in order to avoid the hindering
pedestrian, 
$\vec{v}_\mu^* =\underline{S}_{\beta_{\mu\nu}}^{-1} \vec{v}_\mu$ 
is the inverse rotation to the left side.
\par
In cases, where it is impossible to avoid the individual of type $\nu$
having a velocity $\vec{v}_\nu$, the individual
of type $\mu$ tries to walk with velocity 
$\vec{v}_\mu^* = \vec{v}_\nu$, if $\vec{v}_\nu$ has
a positive component $\vec{v}_\nu \cdot \vec{e}_\mu > 0$ in the
intended direction $\vec{e}_\mu := \vec{v}_\mu^0 / \|\vec{v}_\mu^0 \|$
of motion:
\begin{displaymath}
P_{\mu \nu}^2(\vec{v}_\mu,\vec{v}_\nu;\vec{v}_\mu^*)
= (1-r_\mu)\delta(\vec{v}_\mu^*-\vec{v}_\nu)
\Theta(\vec{v}_\nu\cdot \vec{e}_\mu>0) \, . 
\end{displaymath}
This corresponds to situations, where one moves for a short time
within a gap behind a pedestrian being in the way
(or sometimes, for different directions $\vec{e}_\mu \ne \vec{e}_\nu$, in
front of it).
The {\sl decision function} $\Theta$ is defined by
\begin{displaymath}
 \Theta(x) := \left\{
\begin{array}{ll}
1 & \mbox{if $x$ is fulfilled} \\
0 & \mbox{else.}
\end{array} \right.
\end{displaymath}
If $\vec{v}_\nu\cdot\vec{e}_\mu < 0$ (that means in the case of a 
negative component of the hindering pedestrian's velocity $\vec{v}_\nu$
with respect to the intended direction $\vec{e}_\mu$ 
of movement) it is better for the 
individual(s) to stop ($\vec{v}_\mu^* = \vec{0}$):
\begin{displaymath}
P_{\mu \nu}^3(\vec{v}_\mu,\vec{v}_\nu;\vec{v}_\mu^*)
= (1-r_\mu)\delta(\vec{v}_\mu^*-\vec{0})
\Theta(\vec{v}_\nu\cdot \vec{e}_\mu\le 0) \, . 
\end{displaymath}
This results in\\ \addtocounter{equation}{1}\setcounter{formel4}{\theequation}
\parbox{14.5cm}{\begin{eqnarray*}
& &\frac{1}{m_\nu} \langle \varrho_\mu \rangle \langle \varrho_\nu \rangle
\left[ \chi_{\mu\nu}^k (\vec{v}_\mu^*) - \chi_{\mu\nu}^k(\vec{v}_\mu)\right]
 \\
&\approx& \frac{\langle \varrho_\mu \rangle \langle \varrho_\nu \rangle}
{\tau_{\mu\nu}^*} \cdot \left\{
\begin{array}{ll}
r_\mu[p_{\mu\nu}\langle\underline{S}_{\beta_{\mu\nu}} \vec{v}_\mu \rangle
+ (1 - p_{\mu\nu})\langle \underline{S}_{\beta_{\mu\nu}}^{-1}
\vec{v}_\mu \rangle - \langle \vec{v}_\mu \rangle ]\, ,
& k=1 \\
(1-r_\mu)\langle \Theta_{\mu\nu}\rangle [\langle\vec{v}_\nu \rangle
- \langle \vec{v}_\mu \rangle ] \, , 
& k=2  \\
-(1-r_\mu)(1 - \langle \Theta_{\mu\nu} \rangle )\langle \vec{v}_\mu \rangle
\, , & k=3  
\end{array} \right. \label{momentum} 
\end{eqnarray*}}
\hfill
\parbox{1.3cm}{\begin{displaymath}
\begin{array}{r}
\rule[-2mm]{0cm}{1.1cm}\\
(\theequation.1)\\
(\theequation.2)\\
(\theequation.3)
\end{array}
\end{displaymath}} 
\addtocounter{equation}{1}
\setcounter{formel5}{\theequation}
and \\
\parbox{14.5cm}{\begin{eqnarray*}
& &\frac{1}{m_\nu} \langle \varrho_\mu \rangle \langle \varrho_\nu \rangle
\left[ \chi_{\mu\nu}^k \left(\frac{(v_{\mu,i}^*)^2}{2}\right)
- \chi_{\mu\nu}^k\left(\frac{(v_{\mu,i})^2}{2}\right)\right] \\
&\approx& \frac{\langle \varrho_\mu \rangle \langle \varrho_\nu \rangle}
{\tau_{\mu\nu}^*} \cdot \left\{
\begin{array}{ll}
r_\mu [ p_{\mu\nu} \langle (\underline{S}_{\beta_{\mu\nu}} \vec{v}_\mu)_i^2
\rangle + (1-p_{\mu\nu})\langle (\underline{S}_{\beta_{\mu\nu}}^{-1}
\vec{v}_\mu)_i^2 \rangle - \langle v_{\mu,i}^2 \rangle ] \, ,
& k=1 \\
(1-r_\mu)\langle \Theta_{\mu\nu}\rangle [\langle v_{\nu,i}^2 \rangle
- \langle v_{\mu,i}^2 \rangle ] \, ,
& k=2 \\
-(1-r_\mu)(1 - \langle \Theta_{\mu\nu} \rangle ) \langle v_{\mu,i}^2
\rangle \, ,
& k=3 \, ,
\end{array} \right. \label{energy}
\end{eqnarray*}}
\hfill
\parbox{1.3cm}{\begin{displaymath}
\begin{array}{r}
\rule[-2mm]{0cm}{1.3cm}\\
(\theequation.1)\\
(\theequation.2)\\
(\theequation.3)
\end{array}
\end{displaymath}} 
which allows to calculate (\ref{formel2c}) and (\ref{formel3d})
explicitly.
We have used the abbreviation
\begin{eqnarray}
 \langle \Theta_{\mu\nu} \rangle &=&
 \langle \Theta_{\mu\nu}
 \rangle (\vec{e}_\mu,\langle\vec{v}_\nu \rangle, \theta_\nu)
 := \langle \Theta( \vec{v}_\nu \cdot \vec{e}_\mu > 0 ) \rangle \nonumber \\
& \approx &  \left\{
\begin{array}{ll}
1 - \frac{1}{2}\mbox{e}^{-y_{\mu\nu}^2/(2\theta_\nu)}
+ \frac{y_{\mu\nu}}{\sqrt{2\pi\theta_\nu}} \left[ 1 - \Phi
\left(\frac{y_{\mu\nu}}{\sqrt{2\theta_\nu}}\right)\right] 
& \mbox{if } y_{\mu\nu} \ge 0 \\
\frac{1}{2}\mbox{e}^{-y_{\mu\nu}^2/(2\theta_\nu)} - \frac{|y_{\mu\nu}|}
{\sqrt{2\pi\theta_\nu}} \left[ 1 - \Phi \left( \frac{|y_{\mu\nu}|}
{\sqrt{2\theta_\nu}} \right) \right]
& \mbox{if } y_{\mu\nu} < 0
\end{array} \right. \nonumber
\end{eqnarray}
(see figure \ref{thet}) with
\begin{displaymath}
 y_{\mu\nu}:= \langle \vec{v}_\nu \rangle \cdot \vec{e}_\mu
\end{displaymath}
and the {\sc Gauss}ian {\sl error function}
\begin{displaymath}
 \Phi(z) := \int\limits_0^z \frac{2}{\sqrt{\pi}} \mbox{e}^{-x^2} dx \, .
\end{displaymath}
$(1-r_\mu)(1-\langle \Theta_{\mu\nu}\rangle)$ is the relative
frequency of stopping processes and $r_\mu$ the relative frequency
of avoiding processes to the left or right due to interactions. 
\begin{figure}[htbp]
\unitlength1cm
\begin{picture}(14,11)
\put(0,10.5){$\langle \Theta_{\mu\nu} \rangle$}
\put(13,0){$|y_{\mu\nu}|$}
\end{picture}
\caption[]{The function $\langle \Theta_{\mu \nu} \rangle$ in dependence
of $|y_{\mu \nu}| = |\langle \vec{v}_\nu \rangle \cdot \vec{e}_\mu|$ for the
special case $\theta_{\nu,i}=1$.\label{thet}}
\end{figure}

\subsection{Interpretation} \label{Interpret}

\begin{enumerate}
\item[(a)] {\bf Development of lanes} \\
According to (\arabic{formel4}.1), an asymmetrical avoiding probability
$p_{\mu\nu} \ne 1 - p_{\mu\nu}$ (see \cite{Helbing1}) leads to a 
momentum density that tends to the right (for $p_{\mu\nu} > 1/2$) or
to the left (for $p_{\mu\nu} < 1/2$). This momentum density vanishes when
the products $\langle \varrho_\mu \rangle\langle \varrho_\nu \rangle$ 
of the densities $\langle \varrho_\mu \rangle$, $\langle \varrho_\nu \rangle$
have become zero. Therefore, it causes a separation of different types 
$\mu \ne \nu$ of motion
into several lanes (see figure \ref{separation}). 
This effect can be observed at least for high densities
$\langle \varrho_\mu \rangle$, $\langle \varrho_\nu \rangle$ 
\cite{Helbing1,Soldaten,Verkehrsplanung,Geschwindigkeit}
and has the advantage as well as the purpose 
to reduce the total rate $1/\hat{\tau}_\mu$ of interactions.\\
\begin{figure}[htbp]
\unitlength1cm
\parbox{0.8cm}{\,}\hfill\parbox{15cm}{
\begin{center}
\begin{picture}(10,6)(-0.5,0)
\thicklines
\put(-0.6,-0.2){\line(0,1){6.4}}
\put(3.6,-0.2){\line(0,1){6.4}}
\thinlines
\put(0.5,-0.5){$\Downarrow$}
\put(2.5,6.2){$\Uparrow$}
\put(-1.5,6){(i)}  
\put(0,5.6){\circle{0.4}}
\put(0.4,5.2){\circle{0.4}}
\put(-0.3,5.2){\circle{0.4}}
\put(0.3,4.7){\circle{0.4}}
\put(-0.1,4.3){\circle{0.4}}
\put(0.2,3.9){\circle{0.4}}
\put(0.2,3.9){\vector(1,-2){0.25}}
\put(0,3.4){\circle*{0.4}}
\put(0,3.4){\vector(-1,2){0.25}}
\put(-0.3,3){\circle{0.4}}
\put(0.1,2.5){\circle{0.4}}
\put(-0.2,2){\circle{0.4}}
\put(0.3,1.6){\circle{0.4}}
\put(-0.3,1.3){\circle{0.4}}
\put(0,0.8){\circle{0.4}}
\put(-0.3,0.4){\circle{0.4}}
\put(0.4,0.2){\circle{0.4}}
\put(1.1,0.3){\circle{0.4}}
\put(0.8,0.8){\circle{0.4}}
\put(1.1,1.3){\circle*{0.4}}
\put(1.1,1.3){\vector(1,2){0.25}}
\put(0.9,1.8){\circle{0.4}}
\put(0.9,1.8){\vector(-1,-2){0.25}}
\put(0.8,2.4){\circle{0.4}}
\put(1.5,2.8){\circle*{0.4}}
\put(1.5,2.8){\vector(1,2){0.25}}
\put(0.8,3.2){\circle{0.4}}
\put(1,3.7){\circle{0.4}}
\put(0.7,4.1){\circle{0.4}}
\put(1.3,4.2){\circle{0.4}}
\put(1,4.6){\circle{0.4}}
\put(1.2,5){\circle{0.4}}
\put(0.8,5.4){\circle{0.4}}
\put(1.1,5.9){\circle{0.4}}
\put(1.7,5.6){\circle{0.4}}
\put(1.7,5.6){\vector(-1,-2){0.25}}
\put(2.4,5.5){\circle*{0.4}}
\put(1.8,5.1){\circle*{0.4}}
\put(1.8,5.1){\vector(1,2){0.25}}
\put(2.3,4.8){\circle*{0.4}}
\put(1.9,4.3){\circle*{0.4}}
\put(2.2,3.8){\circle*{0.4}}
\put(1.4,3.4){\circle{0.4}}
\put(1.4,3.4){\vector(-1,-2){0.25}}
\put(2.6,2.3){\circle*{0.4}}
\put(2,2.6){\circle*{0.4}}
\put(1.7,2.2){\circle*{0.4}}
\put(2.3,1.7){\circle*{0.4}}
\put(1.9,1.3){\circle*{0.4}}
\put(2.2,0.9){\circle*{0.4}}
\put(1.5,0.8){\circle*{0.4}}
\put(2,0.4){\circle*{0.4}}
\put(1.8,0){\circle*{0.4}}
\put(2.8,0.2){\circle*{0.4}}
\put(3.2,0.6){\circle*{0.4}}
\put(2.7,1.1){\circle{0.4}}
\put(3,1.6){\circle*{0.4}}
\put(3.3,2){\circle*{0.4}}
\put(2.5,2.9){\circle*{0.4}}
\put(3.2,2.7){\circle*{0.4}}
\put(3.1,3.3){\circle*{0.4}}
\put(2.8,3.8){\circle*{0.4}}
\put(3.2,4.3){\circle*{0.4}}
\put(2.7,4.6){\circle*{0.4}}
\put(3,5.3){\circle*{0.4}}
\put(3.3,5.8){\circle*{0.4}}
\put(5,6){(ii)}
\thinlines
\put(8,1.8){\circle*{0.54}}
\put(8.1,2.1){\vector(1,2){0.9}}
\dashline{0.2}(7.9,2.1)(7,3.83)
\put(7,3.83){\vector(-1,2){0}}
\put(8,4.2){\circle{0.54}}
\put(7.9,3.9){\vector(-1,-2){0.9}}
\dashline{0.2}(8.1,3.9)(9,2.17)
\put(9,2.17){\vector(1,-2){0}}
\put(6.4,2.2){\makebox(0,0){$p_{\nu\mu}$}}
\put(10,2.2){\makebox(0,0){$(1-p_{\nu\mu})$}}
\put(6,3.8){\makebox(0,0){$(1-p_{\mu\nu})$}}
\put(9.6,3.8){\makebox(0,0){$p_{\mu\nu}$}}
\end{picture}
\end{center}
\caption[]{(i) Opposite directions of motion normally use separate lanes.
Avoiding maneuvers are indicated by arrows.
(ii) For pedestrians with an opposite direction of motion it is advantageous,
if both prefer either the right hand side or the left hand side when
trying to pass each other. Otherwise, they would have to stop in order
to avoid a collision. The probability $p_{\mu\nu}$ 
for choosing the right hand side
is usually greater than the probability $(1-p_{\mu\nu})$ 
for choosing the left hand
side.\label{separation}}
}
\end{figure}
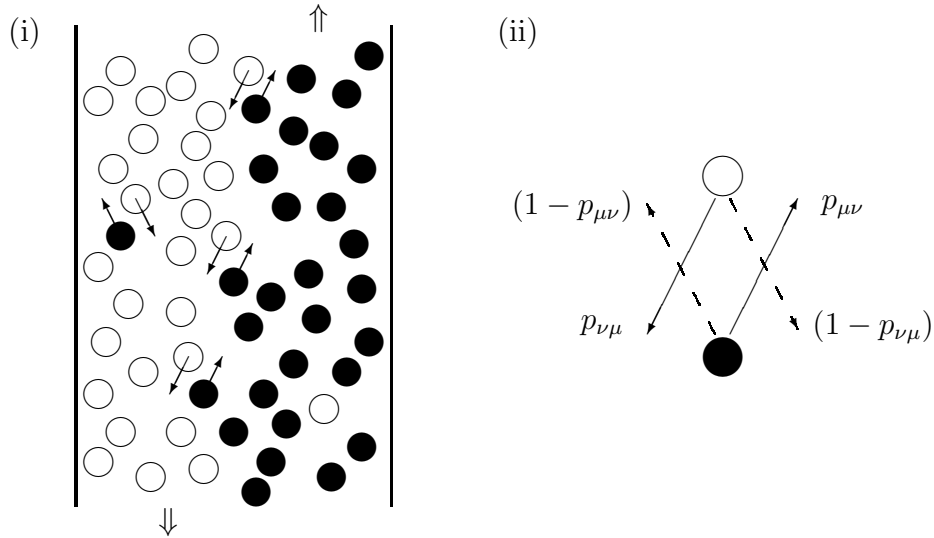
\par
The width of the lanes of two opposing directions 1 and 2
can be calculated from the equilibrium condition
of equal pressure:
\begin{displaymath}
 P_1^e = P_2^e \, , \qquad \mbox{that means} \qquad
\langle \varrho_1 \rangle \theta_1
= \langle \varrho_2 \rangle \theta_2 \, .
\end{displaymath}
Since in the case of a lane of width $W_\mu$ and length $L$
consisting of $N_\mu$ individuals the relation
\begin{displaymath}
\langle \varrho_\mu 
\rangle = m_\mu \langle \rho_\mu \rangle = m_\mu \frac{N_\mu}{W_\mu L}
\end{displaymath}
holds for the mass density $\langle \varrho_\mu \rangle$, we get 
\begin{displaymath}
 \frac{N_1}{N_2}\approx \frac{B_1}{B_2} \, ,
\end{displaymath}
if $m_1\theta_1\approx m_2\theta_2$. That means, the lane width $B_\mu$ will be
proportional to the number $N_\mu$ of individuals (see \cite{Soldaten},
taking into account the {\sc Knudsen} corrections described in section 
\ref{small}.)

\item[(b)] {\bf Crossings} \\
If the direction $\mu$ of motion is crossed by 
the direction $\nu$ of motion, it suffers
a momentum density (\arabic{formel4}.2), 
which causes the individuals of type $\mu$
to be ``pushed'' partly
in direction $\langle \vec{v}_\nu \rangle$ of type $\nu$.
(For the {\sl delay effect} of crossings see \cite{Mayne}.)

\item[(c)] {\bf Pedestrian jams} \\
In order to find out the consequences of (\arabic{formel4}.3)
(that means of stopping processes), we can consider the equation
\begin{equation}
 \frac{\partial \langle \varrho_\mu \rangle\langle \vec{v}_\mu \rangle}
{\partial t} = \frac{1}{\tau_\mu} \langle \varrho_\mu \rangle
(\langle \vec{v}_\mu^0 \rangle - \langle \vec{v}_\mu \rangle )
- \sum_\nu \langle \varrho_\mu \rangle \langle \varrho_\nu \rangle
s_{\mu\nu}^3 \langle \vec{v}_\mu \rangle
\label{stop}
\end{equation}
describing the tendency to walk with the intended velocity $\vec{v}_\mu^0$
as well as stopping processes (see (\ref{formel2b}), (\ref{formel2c})).
\begin{displaymath}
 s_{\mu\nu}^3 := \frac{1 - r_\mu}{\tau_{\mu\nu}^*}(1 - \langle
\Theta_{\mu\nu} \rangle ) 
\end{displaymath}
has been introduced for shortness. 
The stationary solution of equation (\ref{stop}) is
given by
\begin{displaymath}
\langle \vec{v}_\mu \rangle =
\frac{1/\tau_\mu}{1/\tau_\mu 
+ {\displaystyle \sum}_\nu \langle \varrho_\nu \rangle s_{\mu \nu}^3}
\langle \vec{v}_\mu^0 \rangle  =: k_\mu \langle \vec{v}_\mu^0 \rangle 
\end{displaymath}
(compare to \cite{Geschwindigkeit}).
According to (\ref{interrates}) to (\ref{rrr}) we find
\begin{displaymath}
\frac{\partial s_{\mu\nu}^3}{\partial \langle \rho_\nu \rangle}
> 0
\end{displaymath}
and, using the abbreviation $\langle \delta v_{\mu\nu} \rangle
:= \langle \| \vec{v}_\mu - \vec{v}_\nu \| \rangle$,
\begin{equation}
\frac{\partial s_{\mu\nu}^3}{\partial \langle \delta v_{\mu\nu} \rangle}
> 0 \, , \qquad s_{\mu\nu}^3 (\langle \delta v_{\mu\nu} \rangle = 0) = 0 \, .
\label{Stau}
\end{equation}
Due to (\ref{Stau})
a development of pedestrian jams (that means $k_\mu < 1$) is {\sl caused by
the variation} $\delta \vec{v}_{\mu\nu} := \vec{v}_\mu - \vec{v}_\nu$
{\sl of the velocities.} 
This is even the case for 
a lane consisting of individuals of one type $\mu$ only
(where $\langle \varrho_\nu \rangle = 0$
for $\nu \ne \mu$, see (a)), since
$s_{\mu\mu}^3$ is growing with the velocity variance $\theta_\mu$:
\begin{displaymath}
\frac{\partial s_{\mu\mu}^3}{\partial \theta_\mu}
> 0 \, , \qquad s_{\mu\mu}^3 (\theta_\mu = 0) = 0 \, .
\end{displaymath}

\item[(d)] (\arabic{formel5}.3), describes a loss of variance (a loss of
``temperature'')
by stopping processes.
\end{enumerate}

\section{Applications}

\subsection{Optimal motion}

From sections 5.1 and \ref{Interpret} we can 
conclude that the motion of pedestrians can be 
optimized by 
\begin{itemize}
\item avoiding crossings of different 
directions $\mu$ of motion, for example
by bridges or round-about traffic,
\item separation of opposite directions of movement, 
for example by different lanes 
for each direction (the right line being preferred 
\cite{Soldaten,Verkehrsplanung,Geschwindigkeit}),
or walking at a narrow passage {\sl by turns},
\item avoiding great velocity variances $\theta_\mu$, for example by 
walking in formation (as done by the military) \cite{Soldaten},
\item avoidance of obstacles, narrow passages and great densities.
\end{itemize}
These rules are applicable to town- and traffic-planning.

\subsection{Maximal diversity of perceptions}

In a museum or super market the individuals will perceive more details
(and probably buy more goods) if they walk slowly. So the opposite
of the rules in 7.1 could be applied for planning museums and markets.

\subsection{Critical situations}

In critical situations pedestrians may panic. If the mean total
interaction free
time $\hat{\tau}_\mu$ (see (\ref{freetime})) 
is less than the mean reaction time $\zeta_\mu$,
the danger of falling and getting injured is great.
\begin{displaymath}
 \hat{\tau}_\mu > \zeta_\mu
\end{displaymath}
gives a condition for the critical density $\langle \rho_\mu \rangle^{crit}$
of pedestrians that should not be exceeded (see (\ref{interrates}) 
to (\ref{freetime})).

\section{Conclusions}

Starting from the microscopic view of explicit gaskinetic equations
we have derived some fluid dynamic equations for the movement of pedestrians.
These equations are anisotropic 
(that means without local rotational symmetry). They
look similar to the equations for ordinary fluids, but they
are coupled equations for {\sl several} fluids, that means for several
types of motion $\mu$, each consisting
of individuals having approximately the same intended velocity $\vec{v}_\mu^0$.
They also contain
some additional terms that are characteristic 
for {\sl pedestrian} fluids. These
terms are due to the tendency of pedestrians to walk with an intended velocity,
due to pedestrians changing their type (direction) $\mu$ of motion,
and due to interactions between pedestrians (that means due
to avoiding maneuvers). 
\par
For high densities $\langle \rho_\mu \rangle$ the interactions between
pedestrians are very important. As a consequence, a development of
pedestrian jams and of separate lanes for different directions of 
motion is expected. Pedestrian
jams can be understood as an deceleration effect due to avoiding maneuvers,
and become the worse the greater the velocity variance is. The
seperation into several lanes is caused by asymmetrical 
probabilities for avoiding a pedestrian
to the right or to the left. This asymmetry effect
has the advantage to reduce the
situations, where hindering avoiding behavior is necessary.
\par
For pedestrian crowds
the mechanism of approaching equilibrium is essentially
given by the tendency to walk with the intended velocity, not by
interaction processes as in ordinary fluids. As a consequence, the viscosity 
$\eta_\mu$ (that means the coefficient
of internal friction) is growing with the pedestrian density. In addition,
we have seen that variations within pedestrian density will show wave-like
propagation with a velocity $c_\mu$ which depends on the mean reaction 
time.
\par
Last but not least quantities like ``temperature'' and ``pressure'' 
play another role as in ordinary fluids. It can be shown, that 
the ``temperature''
(that means the velocity variance) $\theta_\mu$ 
is produced by the variance of the {\sl intended}
velocities, the individuals {\sl want} to move with. 
As a consequence, two contacting
groups of individuals belonging to different types of motion can show
different ``temperatures''.
This is the case, for example, on a
dance floor. On the other hand, whereas a pressure gradient
is compensating the effect of internal friction in ordinary fluids, 
for pedestrian crowds this role is
played by the accelerating effect of the intended velocity. Therefore a
hyperbolic stationary velocity profile is found instead of a parabolic one. 

\section{Outlook}

Current investigations of pedestrian movement deal with the problem how to
specify the forces $\vec{f}_\mu$, and the rates $\chi_\mu^{..}$ of pedestrians
changing their type of motion. These questions call for a detailled model for
the intentions of pedestrians. 
\par
Pedestrian intentions can be modelled by
stochastic laws. They are functions
\begin{itemize}                                                        
\item of a pedestrian's demand for certain commodities,
\item of the city center entry points (parking lots, metro stations, bus
stops, and so forth),
\item of the location of stores offering the required commodieites,
\item of the expenditures (for example prices, ways), and
\item of unexpected attractions (shop windows, entertainment, and so forth).
\end{itemize}
Models of this kind have been developed and empirically tested
by {\sc Borgers} and 
{\sc Timmermans} \cite{Borg1,Borg2}. A model, which takes into account
pedestrian intentions as well as gaskinetic aspects will be presented
in a forthcoming paper \cite{Helbing3}. It can be formulated in a way that
is also suitable for Monte Carlo simulations of pedestrian dynamics with
a computer. Computer simulations of this kind are an ideal tool for town-
and traffic-planning. Their results can be directly compared with films
of pedestrian crowds (see \cite{Helbing3}).

\paragraph{Acknowledgements} \mbox{ } \\
First I want to thank Prof. Dr. M. R. Schroeder to give me the chance
to work on an interdisciplinary field: modelling the {\sl social}
behavior of pedestrians by {\sl physical} models. Second I am
grateful to Prof. Dr. R. Kree, Prof. Dr. W. Scholl, Prof. Dr. W.
Weidlich, PhD Dr. G. Haag and Dr. R. Reiner 
for their stimulating discussions.
Last but not least I'm obliged to D. Weinmann 
and others for reading, correcting
and commenting on my manuscript.


\end{document}